\documentclass[aps,pre,10pt,twoside,twocolumn,floatfix]{revtex4}

\usepackage{amsmath,amssymb,bm,mathptmx,graphicx}

\usepackage[colorlinks,linkcolor=blue,citecolor=blue,urlcolor=blue]{hyperref}
\usepackage[OT2,T1]{fontenc}
\usepackage[russian,english]{babel}
\DeclareRobustCommand{\cyril}[1]%
{\begingroup\fontfamily{erewhon-TLF}\foreignlanguage{russian}{#1}\endgroup}

\begin{document}

\begin{center}

\end{center}

\title{Density classification performance and ergodicity of the Gacs-Kurdyumov-Levin \\ cellular automaton model IV}

\author{J. Ricardo G. Mendon\c{c}a}
\email[Email: ]{jricardo@usp.br}
\affiliation{\mbox{Escola de Artes, Ci\^{e}ncias e Humanidades, Universidade de S\~{a}o Paulo} \\ \mbox{Rua Arlindo Bettio 1000, Vila Guaraciaba, 03828-000 S\~{a}o Paulo, SP, Brazil}}

\author{Rolf E. O. Sim\~{o}es}
\email[Email: ]{rolf.simoes@inpe.br}
\affiliation{\mbox{Laborat\'{o}rio Associado de Computa\c{c}\~{a}o e Matem\'{a}tica Aplicada, Instituto Nacional de Pesquisas Espaciais} \\ \mbox{Avenida dos Astronautas 1758, Jardim da Granja, 12227-010 S\~{a}o Jos\'{e} dos Campos, SP, Brazil} \\ ${}$}

\received{16 April 2018; published 25 July 2018\\[2ex]}

\begin{abstract}
Almost four decades ago, Gacs, Kurdyumov, and Levin introduced three different cellular automata to investigate whether one-dimensional nonequilibrium interacting particle systems are capable of displaying phase transitions and, as a by-product, introduced the density classification problem (the ability to classify arrays of symbols according to their initial density) in the cellular automata literature. Their model~II became a well known model in theoretical computer science and statistical mechanics. The other two models, however, did not receive much attention. Here we characterize the density classification performance of Gacs, Kurdyumov, and Levin's model~IV, a four-state cellular automaton with three absorbing states---only two of which are attractive---, by numerical simulations. We show that model~IV compares well with its sibling model~II in the density classification task: the additional states slow down the convergence to the majority state but confer a slight advantage in classification performance. We also show that, unexpectedly, initial states diluted in one of the nonclassifiable states are more easily classified. The performance of model~IV under the influence of noise was also investigated and we found signs of an ergodic-nonergodic phase transition at some small finite positive level of noise, although the evidences are not entirely conclusive. We set an upper bound on the critical point for the transition, if any. \\ ${}$ \\
{\noindent}DOI:~\href{https://doi.org/10.1103/PhysRevE.98.012135}{10.1103/PhysRevE.98.012135} 
\end{abstract}

\keywords{Density classification problem, positive probabilities conjecture, ergodic cellular automata, noise, spatially distributed computing}

\maketitle


\section{\label{intro}Introduction}

In 1978, Gacs, Kurdyumov, and Levin (GKL) introduced three different cellular automata (CA), which they called models II, IV, and VI, together with their probabilistic versions (PCA) to investigate whether nonequilibrium interacting particle systems are capable of displaying phase transitions \cite{kurdyumov,gkl}. Their goal was to examine the so-called ``positive probabilities conjecture,'' according to which one-dimensional systems with short-range interactions and positive transition probabilities are always ergodic \cite{dobrushin,petrovskaya,vaserstein,stavskaya,kuznetsov,%
tsirelson,toom,matzak,multi}. This conjecture has been disproved---much to the awe of the practising community---many times since then, with the introduction of several models that have become archetypal models in theoretical computer science and nonequilibrium statistical mechanics \cite{russkiye,holley,%
bennett85,bennett90,grinstein,cuesta,grinjaya,ledoussal,lebowitz,maes,reliable,%
gacs,gray,evans,kafri,bjp30,rakos,paessens}.

As a by-product of their investigations, GKL introduced the density classification problem in the cellular automata literature. The density classification problem consists in classifying arrays of symbols according to their initial density using local rules, and is completed successfully if a correct verdict as to which was the initial majority state is obtained in time at most linear in the size of the input array. Density classification is a nontrivial task for CA in which cells interact over finite neighbourhoods, because then the cells have to achieve a global consensus cooperating locally only. Ultimately, that means that information should flow through the entire system, be processed by the cells, and be not destroyed or become incoherent in the process---entropy must lose to work, a relevant property in the theoretical analysis of data processing and storage under noise \cite{bennett85,bennett90,grinstein}. For one-dimensional locally interacting systems of autonomous and memoryless cells, emergence of collective behavior is required in these cases. In this context, GKL model II has been extensively scrutinized as a model system related with the concepts of emergence, communication, efficiency, and connectivity \cite{mitchell93,crutch,landbelew,%
fuks,sipcapron,juille,boolean,amaral,mitchell05,moreira,briceno,stone,regnault,%
taati15,stavs,assembly,Wolnik2017}. The search for efficient density classifiers is an entire subfield in the theory of cellular automata. The usual candidates are the so-called eroders, a class of CA to which the GKL models belong (see Sec.~\ref{model}). Unfortunately, the general problem of defining or discerning eroders is algorithmically unsolvable \cite{russkiye,reliable,gacs,gray}. As such, the proposition and analysis of particular models has always been carried out with great interest. Current trends, advances and open problems related with the density classification problem are reviewed in \cite{fates,ppbortot,ppreview}.

In this paper we characterize the density classification performance of Gacs, Kurdyumov, and Levin's model~IV, a four-state cellular automaton with three absorbing states, by numerical simulations. To our knowledge, the model never received a thorough examination of its basic dynamics and properties since its proposition. We show that GKL model~IV compares well with its sibling model~II in the density classification task, although it takes longer to converge to the right answer. We also investigate the performance of model~IV under the influence of noise and show that, most likely, it displays an ergodic-nonergodic transition at some finite small level of noise, although the evidences are not conclusive.

The paper goes as follows: in Section~\ref{model} we introduce the GKL model IV, describe its transition rules, and discuss some of its properties. In Section~\ref{performance} we describe our numerical simulations and discuss the density classification performance of the model in its deterministic version, inlcuding a comparison with GKL model II, while in Section~\ref{noisy} we examine the behavior of the model under the influence of noise. In Section~\ref{summary} we summarize our findings and discuss our results. An appendix displays the complete rule table of GKL model~IV.


\section{\label{model}Gacs, Kurdyumov, and Levin's CA model IV}

Gacs, Kurdyumov, and Levin's model IV (GKL-IV for short) is a four-state CA with state space given by $\Omega_{\Lambda} = \{{\to},{\leftarrow},{\uparrow},{\downarrow}\}^{\Lambda}$, with $\Lambda \subseteq \mathbb{Z}$ a finite array of $|\Lambda|=L \geq 1$ cells under periodic boundary conditions, and transition function $\Phi_{\rm IV}: \Omega_{\Lambda} \to \Omega_{\Lambda}$ that given the state $\bm{x}^{t} = (x_{1}^{t}, x_{2}^{t}, \ldots, x_{L}^{t})$ of the CA at instant $t$ determines the state $x_{i}^{t+1} = [\Phi_{\rm IV}(\bm{x}^{t})]_{i} = \phi_{\rm IV}(x^{t}_{i-1}, x^{t}_{i}, x^{t}_{i+1})$ of the CA at instant $t+1$ by the rules
\begin{widetext}
\begin{subequations}
\label{gkl-iv-rules}
\begin{align}
\label{gkl-iv-1}
\phi_{\rm IV}({\to},\, x_{i},\, x_{i+1}) &= {\to}, \mkern30.5mu\textrm{if }
x_{i},\,x_{i+1} \ne {\leftarrow}, \\
\label{gkl-iv-2}
\phi_{\rm IV}(x_{i-1},\, {\to},\, x_{i+1}) &=
\begin{cases}
  {\downarrow}, & \textrm{if } x_{i-1} \in \{{\leftarrow},{\uparrow}\}, \\
  {\to},        & \textrm{otherwise},
\end{cases} \\
\label{gkl-iv-3}
\phi_{\rm IV}(x_{i-1},\, x_{i},\, x_{i+1}) &= {\uparrow}, \mkern39.5mu\textrm{if }
x_{i} \in \{{\uparrow},{\downarrow}\} \textrm{ and rule (\ref{gkl-iv-1}) does not apply}.
\end{align}
\end{subequations}
\end{widetext}
Rules (\ref{gkl-iv-1})--(\ref{gkl-iv-3}) are redundant---for example, transitions $\phi_{\rm IV}({\to},{\to},\{{\to},{\uparrow},{\downarrow}\})$ are defined both by rules (\ref{gkl-iv-1}) and (\ref{gkl-iv-2})---and incomplete, since they define only $42$ of the $64$ possible transitions. For example, they do not define the important transition $\phi_{\rm IV}({\leftarrow},{\leftarrow},{\to})$, see Figure~\ref{fig:miss}. The missing transitions are determined by the supplemental reflection rule
\begin{equation}
\label{gkl-iv-r}
\phi_{\rm IV}(x_{i-1}, x_{i}, x_{i+1}) =
\phi_{\rm IV}(x^{*}_{i+1}, x^{*}_{i}, x^{*}_{i-1})^{*},
\end{equation}
with ${\to}^{*} = {\leftarrow}$, ${\leftarrow}^{*} = {\to}$, ${\uparrow}^{*} = {\uparrow}$, and ${\downarrow}^{*} = {\downarrow}$. The reflection rule supplements rules (\ref{gkl-iv-1})--(\ref{gkl-iv-3}) in their order of appearance and does not substitute a transition that has already been defined. Since the GKL-IV rules are somewhat unwieldy, we give the complete rule table of the CA in the appendix.

\begin{figure}[t]
\centering
\includegraphics[viewport=50 35 740 580, scale=0.30, clip]{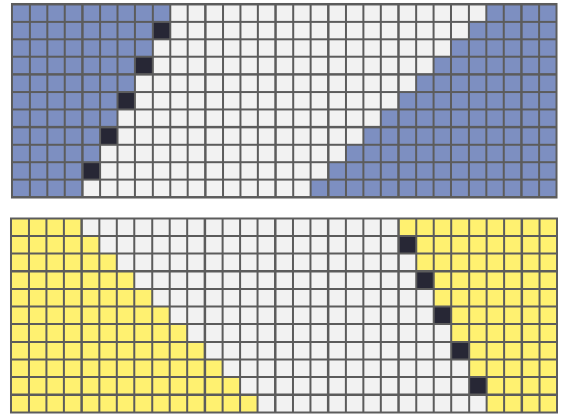}
\caption{\label{fig:gkl-iv}Dynamics of GKL-IV in two schematic situations exemplifying the eroder mechanism. Time flows downward. Cells are color-coded as yellow $\equiv {\to}$, purple $\equiv {\leftarrow}$, white $\equiv {\uparrow}$, and black $\equiv {\downarrow}$. Note the difference in the propagation speed of the left and right fronts intermediated by the ${\downarrow}$ arrows. The upper configuration will clearly converge to the all ${\leftarrow}$ (purple) state, while the lower configuration will converge to the all ${\to}$ (yellow) state.}
\end{figure}

The rationale behind the GKL-IV rules is that of an ``eroder.'' An eroder CA is capable of erasing ``errors'' in the initial configuration, which for a density classifier means to erase the symbols of the minority phases. In GKL-IV, this is achieved by the propagation of state ${\to}$ over states ${\uparrow}$ and ${\downarrow}$ from the left, rule (\ref{gkl-iv-1}), and of state ${\leftarrow}$ over states ${\uparrow}$ and ${\downarrow}$ from the right, reflection (\ref{gkl-iv-r}) of rule (\ref{gkl-iv-1}). Rules (\ref{gkl-iv-2}) and its mirrored symmetric rule along with rule (\ref{gkl-iv-3}) define two other processes. The first consists in the annihilation of states ${\leftarrow}$ and ${\to}$ when they are adjacent; the second process consists in the propagation of state ${\uparrow}$ over states ${\to}$ and ${\leftarrow}$ from the right and left, respectively. These processes of annihilation and propagation occur by substitution, as they are intermediated by state ${\downarrow}$, which is then converted to state ${\leftarrow}$ in the next time step, leading to the continued propagation of state ${\uparrow}$. Overall, these rules promote the propagation of state ${\uparrow}$ over states ${\to}$ and ${\leftarrow}$ at half the speed of the inverse propagation of states ${\to}$ and ${\leftarrow}$ over state ${\uparrow}$, because of the intermediate step involving state ${\downarrow}$. The only role played by state ${\downarrow}$ in the dynamics of GKL-IV is that of delaying the conversion of states ${\to}$ and ${\leftarrow}$ into state ${\uparrow}$ such that the CA can erode states ${\uparrow}$ and ${\downarrow}$ within islands of the minority phase towards the stationary configuration of the majority state. Figure~\ref{fig:gkl-iv} displays the eroder mechanism in action in two schematic situations.

In \cite{gkl}, the authors state that the states ${\to}$ and ${\leftarrow}$ are attracting states for models II and IV, further noting that ``evidently [models II and IV] do not have other attracting states.'' It happens, however, that GKL-IV has three absorbing states, as it can be seen from the transitions $\phi({\to},{\to},{\to}) = {\to}$, $\phi({\leftarrow},{\leftarrow},{\leftarrow}) = {\leftarrow}$, and $\phi({\uparrow},{\uparrow},{\uparrow}) = {\uparrow}$. That the states $({\to},{\to},\ldots,{\to})$ and $({\leftarrow},{\leftarrow},\ldots,{\leftarrow})$ are attracting is a theorem of GKL \cite{gkl}, revisited in \cite{maes}. The state $({\uparrow},{\uparrow},\ldots,{\uparrow})$, despite being absorbing, may not be attracting, since it may not be true that if we disturb it in finitely many places it will recur in finite time. In our simulations on relatively small systems, however, we observed the convergence of the GKL-IV CA to the state $({\uparrow},{\uparrow},\ldots,{\uparrow})$ many times. Rough initial estimates indicated that for random, uncorrelated initial configurations in which cells initially get one of the four possible states with equal probabilities, the final configuration converges to $({\uparrow},{\uparrow},\ldots,{\uparrow})$ about $1\%$ of the times. We thus asked whether GKL-IV can classify initial configurations with majority of cells in state ${\uparrow}$, even if it was not designed for the task. As we will see in Section~\ref{performance} the answer is nearly never.


\section{\label{performance}GKL-IV density classification performance}

\subsection{\label{iv_vs_ii}GKL-IV \textit{vs}. GKL-II performance}

\begin{figure*}[t]
\centering
\includegraphics[viewport=10 0 530 430, scale=0.40, clip]{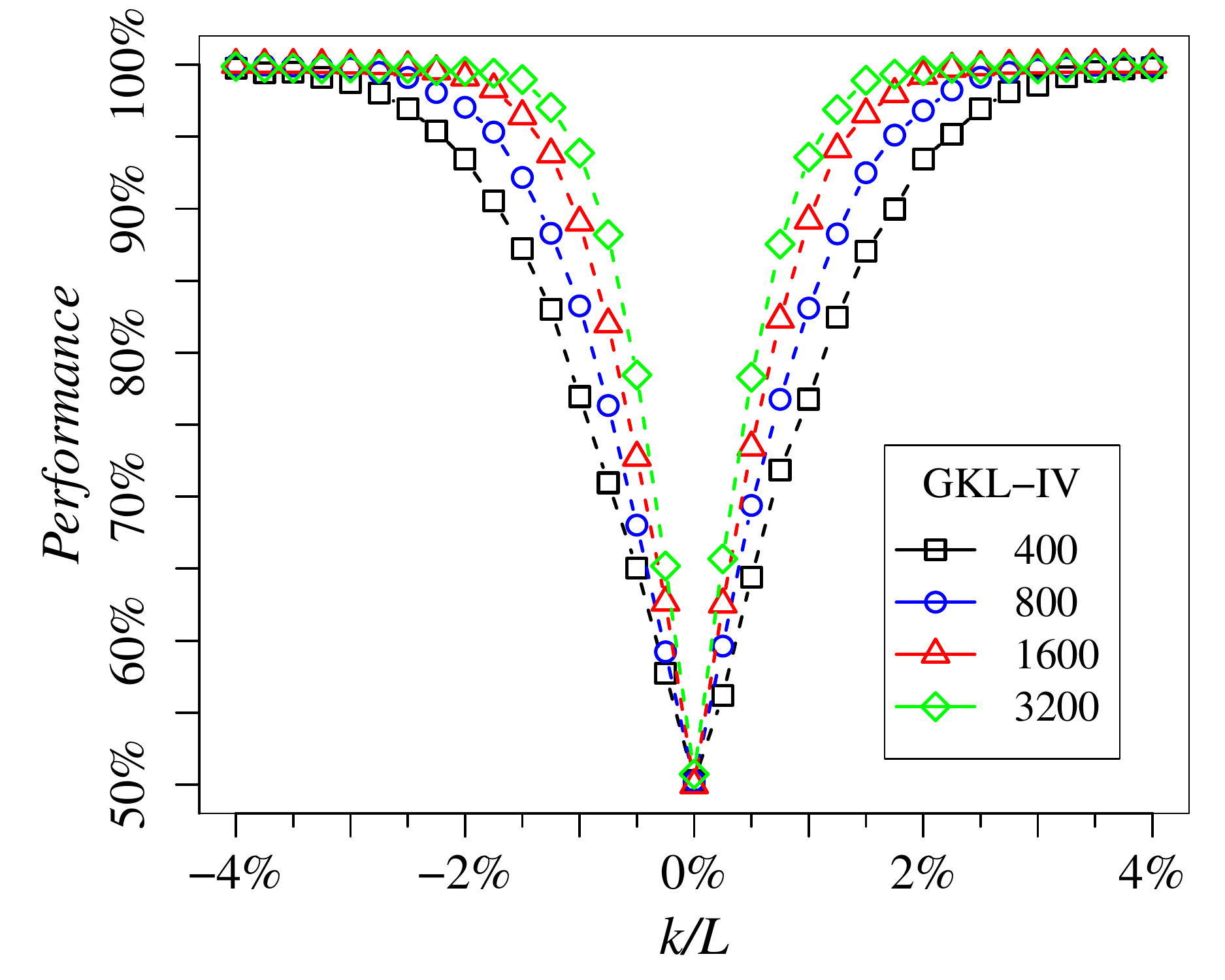}
\hspace{2em}
\includegraphics[viewport=0 0 530 430, scale=0.40, clip]{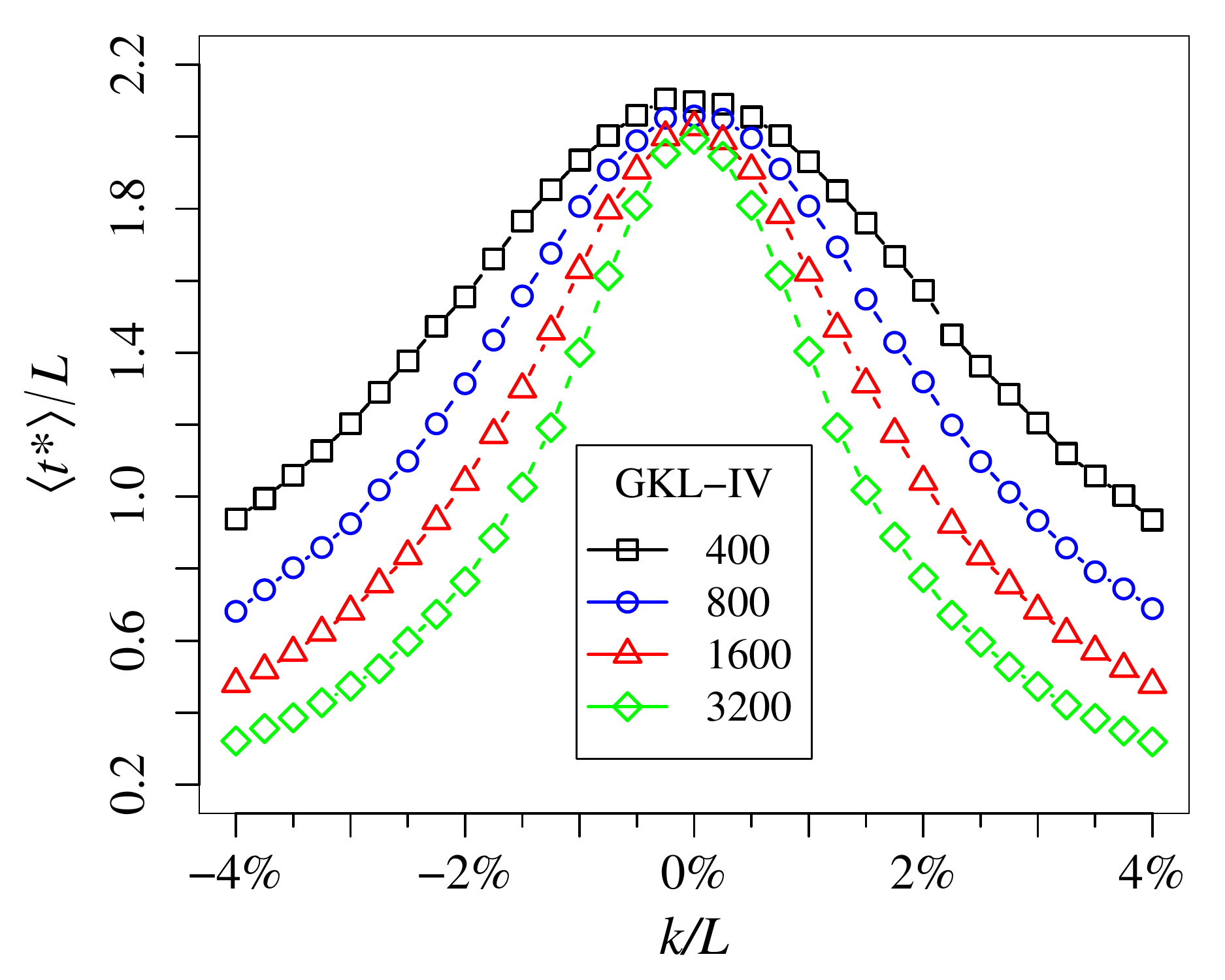}
\\
\includegraphics[viewport=10 0 530 430, scale=0.40, clip]{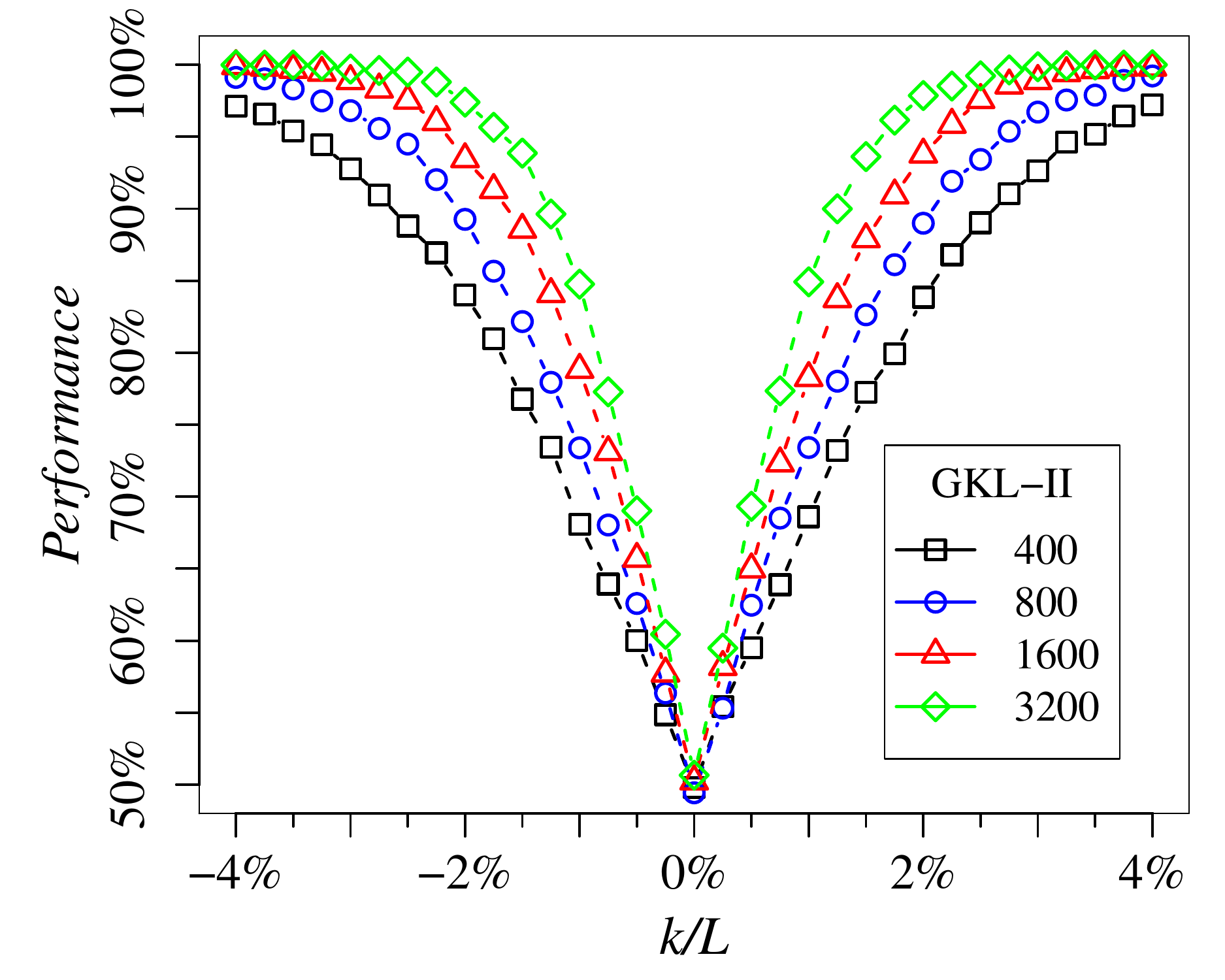}
\hspace{2em}
\includegraphics[viewport=0 0 530 430, scale=0.40, clip]{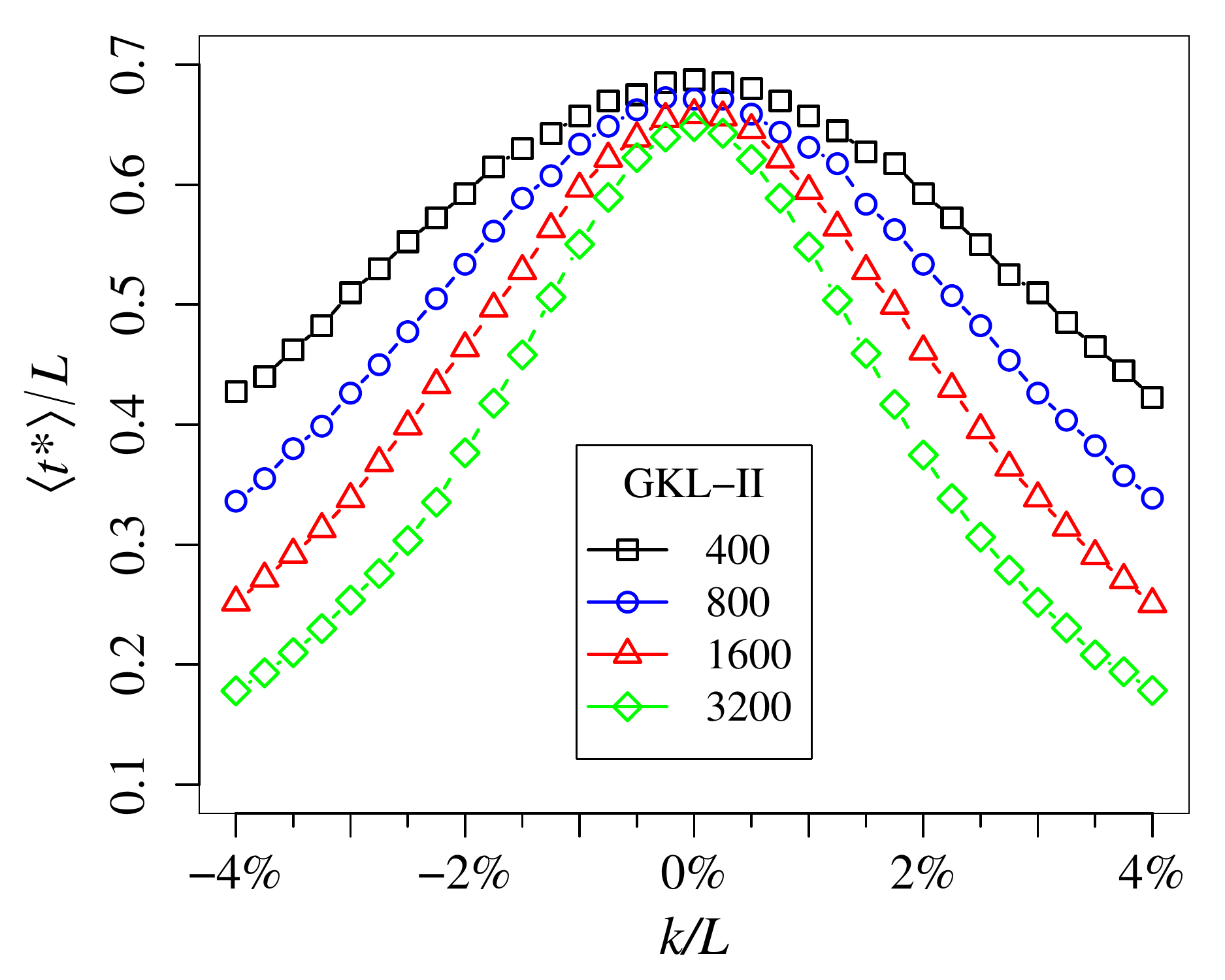}
\caption{\label{fig:performance}Density classification performance of GKL-IV and GKL-II, shown for comparison, for CA of lengths $L=400$, $800$, $1600$, and $3200$. Parameter $k=\frac{1}{2}(N_{{\to}}-N_{{\leftarrow}})$ represents the imbalance between the number of cells in states ${\to}$ and ${\leftarrow}$ in the initial configuration. Left panels: neither GKL-II nor GKL-IV is a perfect classifier, but GKL-IV performs better. Right panels: average time $\langle t^{*} \rangle$ to converge to the stationary state in units of the length of the CA. In this case, GKL-IV performs worse than GKL-II. Data correspond to the performance measured over $10\,000$ random initial configurations for each given $k/L$; error bars ($\pm 0.5\%$ or less) are of the order of the sizes of the symbols shown.}
\end{figure*}

Let $N_{s}$ be the number of cells in the state $s \in \{{\to}, {\leftarrow}, {\uparrow}, {\downarrow}\}$, with $N_{{\to}}+N_{{\leftarrow}}+N_{{\uparrow}}+N_{{\downarrow}}=L$, the size of the array. Given an initial assignment of the numbers $N_{s}$, the main observables of the CA are the empirical time-dependent number of cells in state $s$ given by
\begin{equation}
\label{density}
N_{s}(t) = \sum_{i=1}^{L} \delta(x^{t}_{i},s),
\end{equation}
where $\delta(\cdot,\cdot)$ is the Kronecker delta symbol. GKL argued in \cite{gkl} that in the stationary state either the state ${\to}$ or the state ${\leftarrow}$ completely dominates the CA, with the dominance depending on which state, ${\to}$ or ${\leftarrow}$, respectively, prevails in the initial configuration; states ${\uparrow}$ and ${\downarrow}$ are washed out by the dynamics and do not survive to the stationary state.

We assess the density classification performance of the GKL-IV CA by direct simulations as follows. We set the array size $L$ to a multiple of $4$ and then assign $L/4$ randomly chosen cells to each of the states ${\uparrow}$ and ${\downarrow}$, $L/4+k$ cells to state ${\to}$, and $L/4-k$ cells to state ${\leftarrow}$, with $k$ an integer parameter that can be varied in the range $-L/4 \leq k \leq L/4$. If GKL-IV can classify density, we expect that when $k>0$ the stationary state will be the all ${\to}$ state while when $k<0$ the stationary state will be the all ${\leftarrow}$ state. We then evolve the CA array and track the empirical densities until at some time $t^{*}$ either $N_{{\to}}(t^{*})=L$ or $N_{{\leftarrow}}(t^{*})=L$. If initially $k>0$ ($k<0$) and the stationary state becomes the all ${\to}$ (respectively, all ${\leftarrow}$) state, then GKL-IV has classified the initial state succesfully, otherwise it has failed. We also consider that the CA failed if after $4L$ time steps the array did not converge to one of those two states, but this did not happen in our simulations. When $k=0$, we compute the performance of the GKL-IV array as the number of times that it converges to the all ${\to}$ state. The choice between the all ${\to}$ or the all ${\leftarrow}$ states in this case is irrelevant because the GKL-IV rules are reflection-symmetric with respect to these states and our array is periodic; see Table~\ref{tab:gkl_iv_rules}.
For each given $k$, the performance of the CA is estimated as the fraction $\hat{p}$ of correct classifications measured over $n=10\,000$ random initial configurations, with standard deviation estimated as $\sqrt{\hat{p}(1-\hat{p})/n}$. The results appear in Figure~\ref{fig:performance} for CA of lengths $L=400$, $800$, $1600$, and $3200$. At the ``worst case'' of $k/L=0$ for a CA of $L=400$ cells we measured $\langle n \rangle = 0.503 \pm 0.005$.


For comparison, results for the sibling GKL model II (GKL-II for short) are also displayed in Figure~\ref{fig:performance}. GKL-II is a two-state CA that evolves by the following rule \cite{gkl,maes}: if the state $x_{i}^{t}$ of the cell at instant $t$ is $+1$ (or, equivalently, ${\to}$), then at instant $t+1$ it takes the state of the majority state of itself and the first and the third neighbors to its right, otherwise, if the state of the cell at instant $t$ is $-1$ (or ${\leftarrow}$), it takes at instant $t+1$ the state of the majority state of the same neighborhood but in the opposite direction. In symbols, $x_{i}^{t+1} = [\Phi_{\rm II}(\bm{x}^{t})]_{i} = \phi_{\rm II}(x_{i}^{t},\, x_{i+s}^{t},\, x_{i+3s}^{t})$ with
\begin{equation}
\phi_{\rm II}(x_{i}^{t},\, x_{i+s}^{t},\, x_{i+3s}^{t}) =
\mathrm{maj}(x_{i}^{t},\, x_{i+s}^{t},\, x_{i+3s}^{t})
\end{equation}
and $s = x_{i}^{t} = \pm 1$. The GKL-II rule is not nearest-neighbor but observes a generalization of the reflection rule (\ref{gkl-iv-r}), to wit, $\phi_{\rm II}(x_{i},\, x_{j},\, x_{k})=\phi_{\rm II}(x_{\sigma(i)}^{*},\, x_{\sigma(j)}^{*},\, x_{\sigma(k)}^{*})^{*}$, where $x^{*}=-x$ and $\sigma$ is any permutation of \mbox{$i$, $j$, $k$}. The GKL-II automaton achieves $81.6\%$ performance in a test consisting of classifying something between $10^4$ and $10^7$ random initial configurations of an array of $L=149$ (sometimes also $599$ and $999$) cells close to the ``critical density'' $N_{\to}=1/2$ \cite{crutch,sipcapron}. Improvements of the GKL-II rule by humans as well as by genetic and co-evolution programming techniques were able to upgrade the success rate of GKL-II (under the same evaluation protocol) to $86.0\%$ \cite{sipcapron,juille}. There are other figures published for the GKL-II and other CA (see, e.\,g., \cite{landbelew}), but the reader must be aware that the direct comparison of CA performance numbers is not straighforward because of the use of different array lengths, sets of initial configurations, and evaluation/measurement protocols. Some of these numbers and issues are reviewed in \cite{fates,ppbortot,ppreview}.


Note that, since GKL-IV has $4$ states instead of the $2$ states of GKL-II, one might argue that the proper quantity to be used in the comparison of the two CA would be relative imbalance between the classifiable states only, which in our case would read $k/(\frac{2}{4}L)=2k/L$. We have adopted, however, the point of view of a ``client application'' that wants to classify the majority between two possible states with a CA. From this point of view, it does not matter if the CA has $2$ or more states.

We see from Figure~\ref{fig:performance} that GKL-IV is not a perfect classifier: for random initial configurations, sometimes it converges to the wrong answer. Otherwise, when the imbalance $k/L$ between the ${\to}$ and the ${\leftarrow}$ states in the initial state is larger than $\sim 2\%$, the GKL-IV classification performace exceeds $95\%$, an excellent result. This performance is considerably better than the one for GKL-II, that at $k/L=2\%$ is only about $\sim 85\%$ and reaches the $95\%$ mark only for $k/L \gtrsim 3\%$.
It should be remarked that the density classification task cannot be achieved without misclassifications by any single locally interacting two-state cellular automaton. Indeed, under the requirement that all the cells of the automaton must converge to the same state as the majority state in the initial configuration, no automata can achieve 100\% efficiency \cite{landbelew,sipcapron}. These results establish the need for probabilistic rules and imperfect quality measurements.

The panels on the right in Figure~\ref{fig:performance} display the time needed to converge to the stationary state as a function of $k/L$. We see that as the imbalance becomes smaller the time to converge grows, but it never grows more than linearly with the size of the array. Even for large instances of the problem (large $L$) in the difficult region $k/L \ll 1$, GKL-IV converges fast to the solution. In this regard, however, GKL-II exceeds GKL-IV almost by a factor of $3$. It seems that the additional states of GKL-IV provide more ``error-correction,'' while at the same time retarding the convergence to the majority state.


\subsection{\label{upupup}Influence of the nonattractive state $({\uparrow},{\uparrow},\ldots,{\uparrow})$}

\begin{figure}[t]
\centering
\includegraphics[viewport=10 0 535 430, scale=0.40, clip]{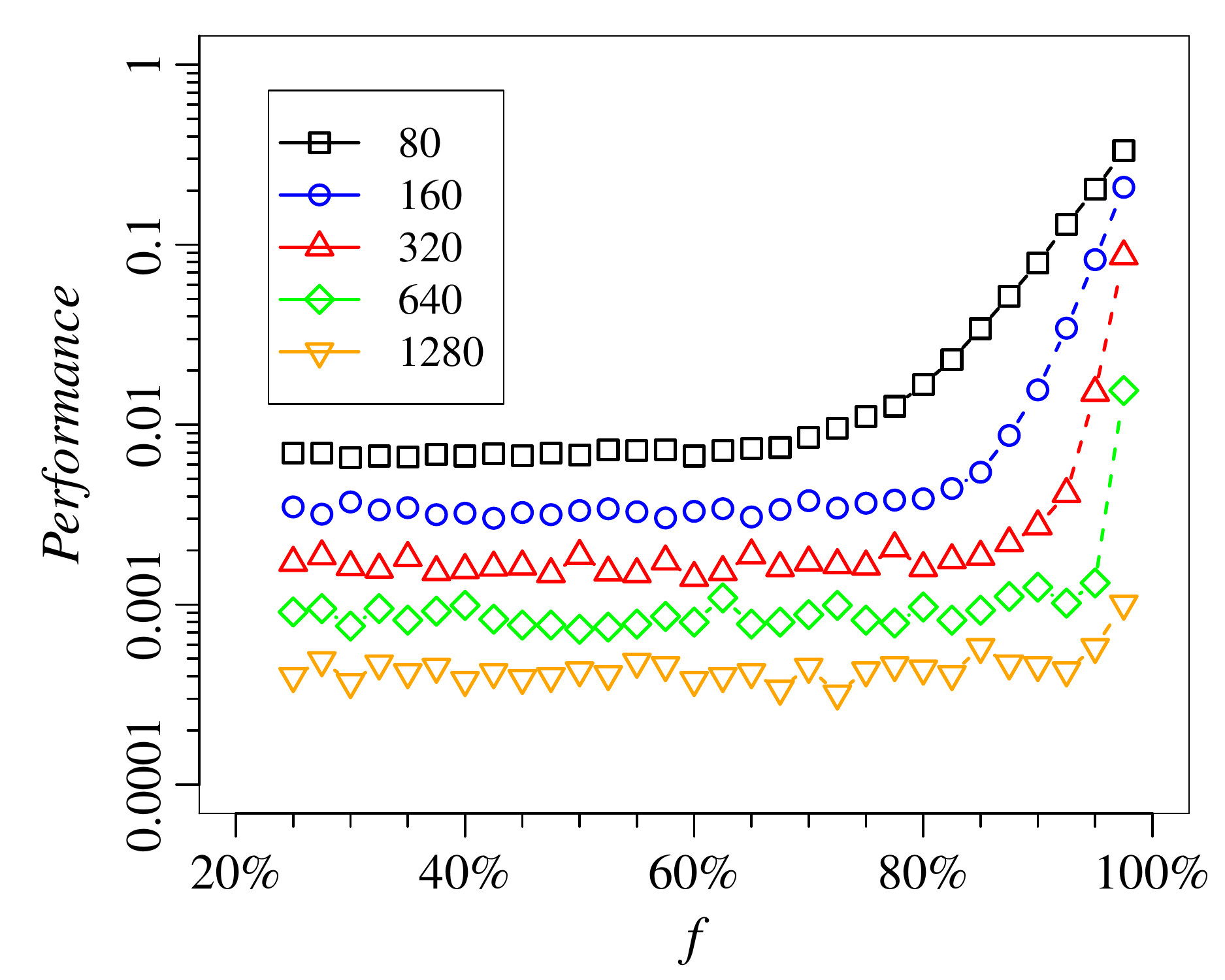}
\caption{\label{fig:ups}Density classification performance of GKL-IV with respect to the ${\uparrow}$ state for CA of lengths $80 \leq L \leq 1280$ and fraction $0.25 \leq f \leq 0.975$ of ${\uparrow}$ arrows in the initial state. Each point corresponds to an average over $100\,000$ random initial configurations; error bars are of the order of the symbols shown or smaller (note the logarithmic scale). The stationary state seldom converges to the absorbing state of all ${\uparrow}$ states but for the smallest arrays.}
\end{figure}

We found that GKL-IV sometimes converges to the all ${\uparrow}$ state, which is also one of its absorbing configurations. For relatively small array sizes and random initial configurations in which each cell starts in one of the four possible states with equal probability $1/4$, we observed that the array converges to the state $({\uparrow},{\uparrow},\ldots,{\uparrow})$ approximately $1\%$ of the times. To quantify this behavior, we performed the following numerical experiment: we initially assign a fraction $\frac{1}{4} \leq f \leq 1$ of the $L$ cells to state ${\uparrow}$ and distribute the remaining $(1-f)L$ cells randomly to the other three possible states---such that $N_{{\uparrow}}(0)=fL$ exactly and, on average, $N_{s}(0)=\frac{1}{3}(1-f)L$ for each of the other possible states---, evolve the dynamics and observe the approach to stationarity. The results are summarized in Figure~\ref{fig:ups}. As we can see from that figure, even with as much as $95\%$ of the cells initially in the state ${\uparrow}$ GKL-IV cannot really classify initial states with majority of cells in the state ${\uparrow}$ except for the smallest arrays---and even in these cases, only very badly, at a rate of $\sim 10\%$. Data suggest that as $L \nearrow \infty$ the stationary state is never $({\uparrow},{\uparrow},\ldots,{\uparrow})$ unless $f=1$ exactly. The occasional convergence to the stationary state of all  ${\uparrow}$ arrows is thus a feature of finite small size systems.




\begin{figure}[t]
\centering
\includegraphics[viewport=10 0 535 430, scale=0.40, clip]{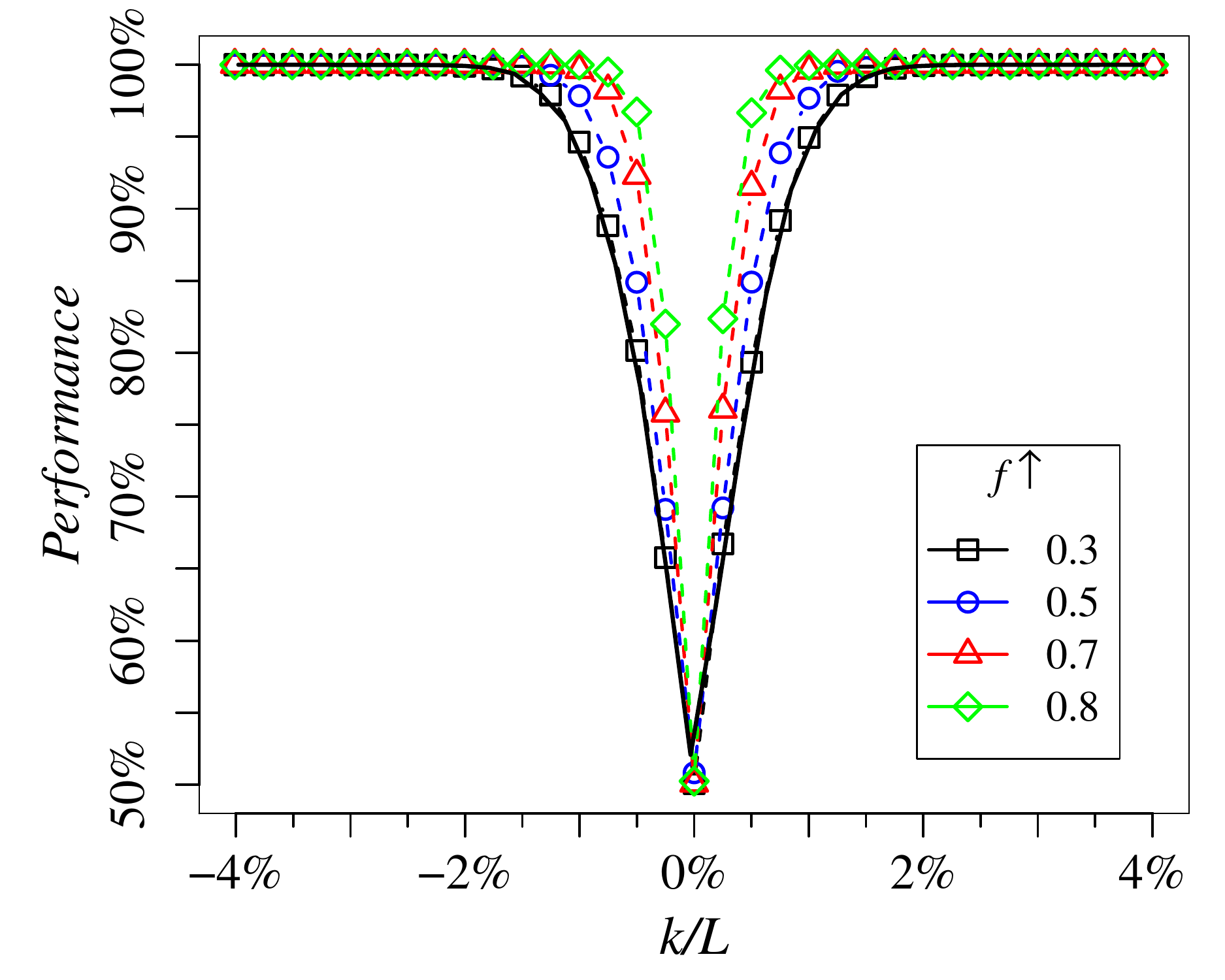} \\
\includegraphics[viewport=10 0 535 430, scale=0.40, clip]{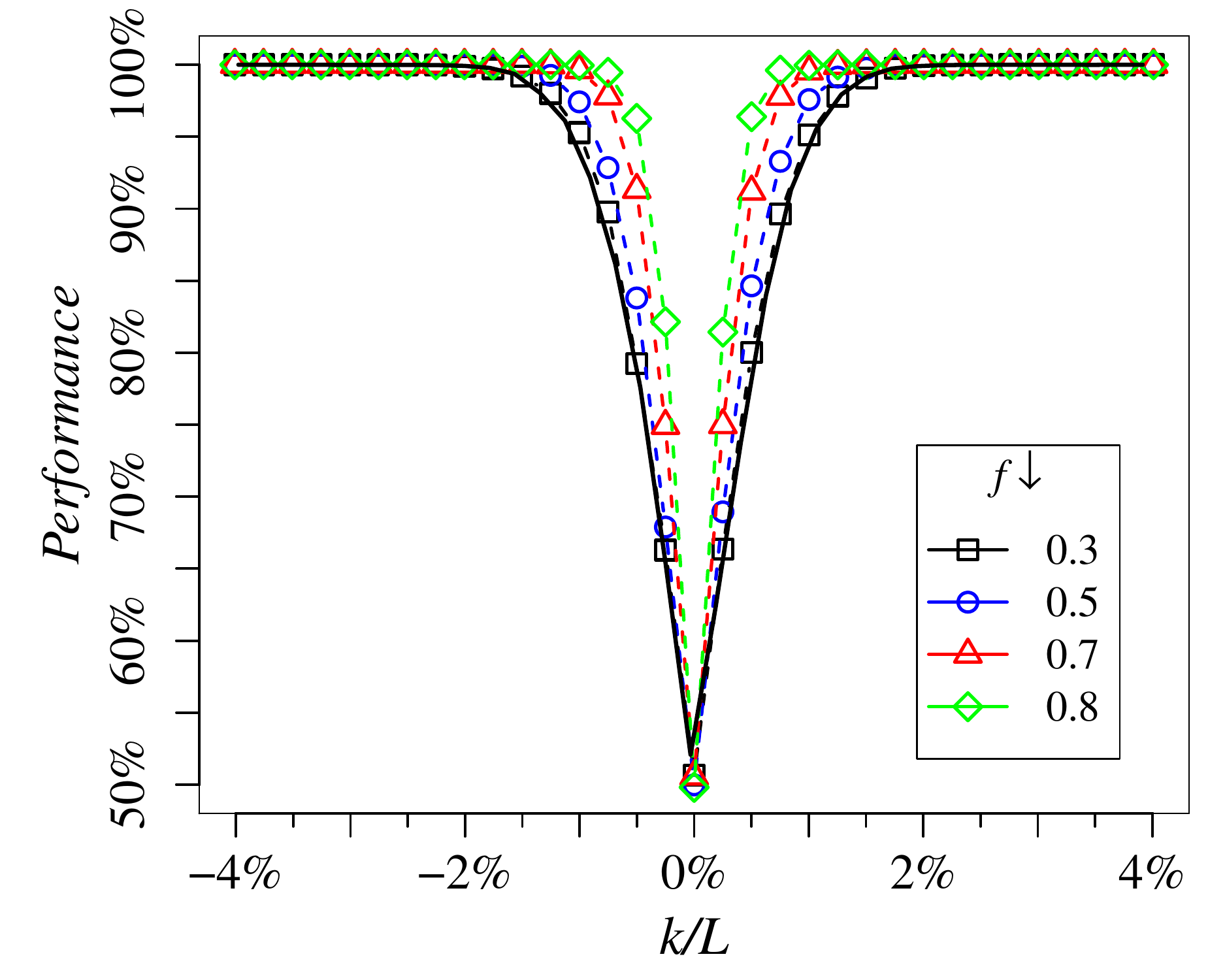}
\caption{\label{fig:fk}Density classification performance of GKL-IV in the presence of a fraction $f$ of ${\uparrow}$ (upper panel) or ${\downarrow}$ (lower panel) states in the initial state for a CA of $3192$ cells. Parameter $k=\frac{1}{2}(N_{{\to}}-N_{{\leftarrow}})$ represents the imbalance between the number of cells in states ${\to}$ and ${\leftarrow}$ in the initial configuration. Performance increases with increasing $f$. The thick solid lines display the performance of GKL-IV with $N_{{\uparrow}}=\frac{1}{4}L$ (resp.~$N_{{\downarrow}}=\frac{1}{4}L$), i.\,e., $f=0.25$. Data correspond to average performance over $10\,000$ random initial configurations for each given $f$ and $k/L$; error bars ($\pm 0.5\%$ or less) are of the order of the sizes of the symbols shown.}
\end{figure}

\begin{figure}[t]
\centering
\includegraphics[viewport=140 140 660 480, scale=0.40, clip]{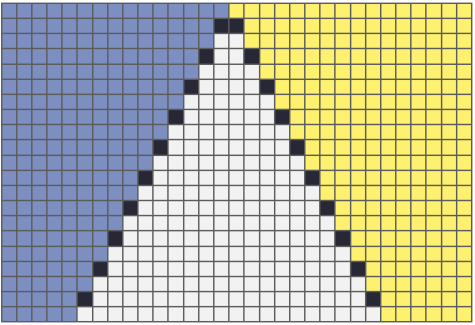}
\caption{\label{fig:miss}Schematic configuration displaying a domain wall ${\leftarrow}\,{\leftarrow}\,{\to}$ (or its reflection ${\leftarrow}\,{\to}\,{\to}$), that may lead to missclassifications in the long run. Cells are color-coded as in Figure~\ref{fig:gkl-iv}. Higher densities of the otherwise inocuous state ${\uparrow}$ (white) dilute such domain walls, slightly improving the density classification performance of GKL-IV.}
\end{figure}

A closely related question is whether the density of states $\uparrow$ (and perhaps $\downarrow$) impacts the classification performance of GKL-IV. We assess this impact by measuring the classification performance of GKL-IV as follows: given an initial assignment of $fL$ of cells in state ${\uparrow}$, the remaining $(1-f)L$ cells are divided between $\frac{1}{3}(1-f)L$ cells in state ${\downarrow}$, $\frac{1}{3}(1-f)L+k$ cells in state ${\to}$ and $\frac{1}{3}(1-f)L-k$ cells in state ${\leftarrow}$, with $\frac{1}{4} \leq f \leq 1$ such that initially $N_{{\uparrow}} \geq N_{{\downarrow}}$. We then measure the performance of the CA in terms of $f$ and $k/L$ for a fixed $L=3192$ (which is close to $3200$ but is divisible by $3$ and $4$). The results appear in Figure~\ref{fig:fk}. We see that the impact of the number of ${\uparrow}$ in the initial state is noticeable, with an unexpected improvement in the density classification performance of GKL-IV at higher densities of ${\uparrow}$ arrows in the initial state near the ``critical density'' $k/L=0$. For example, at $k/L=0.25\%$, the performance of GKL-IV increases from $\sim 67\%$ at $f=0.3$ to $\sim 82\%$ at $f=0.8$. As the imbalance $k/L$ becomes larger, the performance gain tapers off, but remains measurable. The same phenomenon was observed when the initial state is diluted in ${\downarrow}$ arrows. In both cases, the performance remains larger than the one measured by the protocol of Section~\ref{iv_vs_ii}, which corresponds to $N_{{\uparrow}}=\frac{1}{4}L$ (i.\,e., $f=0.25$), displayed in Figure~\ref{fig:fk} as thick solid black lines. One possible explanation for this improvement is that upon dilution in a field of ${\uparrow}$ or ${\downarrow}$ arrows, extended regions of ${\leftarrow}$ and ${\to}$ states separated by domain walls of the type ${\leftarrow}\,{\leftarrow}\,{\to}$ or ${\to}\,{\to}\,{\leftarrow}$\,---configurations that may lead to missclassifications in the long run, see Figure~\ref{fig:miss}---hardly form. Other GKL-IV processes certainly play a role in this behavior, although their identification is not immediate.



\section{\label{noisy}The GKL-IV model under noise}

A CA with rules depending on a random variable becomes a probabilistic CA (PCA) \cite{russkiye}. Let us denote the probabilistic transition function of the GKL-IV model under noise by $\Phi_{\rm IV}^{(\alpha)}$, where the real parameter $\alpha \in [0,1]$ denotes the level of noise imposed to the dynamics. If $\alpha = 0$, GKL-IV becomes the deterministic CA given by transition rules (\ref{gkl-iv-rules})--(\ref{gkl-iv-r}), otherwise with probability $\alpha > 0$ the transition rules fail in some specific manner, leading to an evolved state that may be at variance with the one prescribed by the deterministic transition rules. In their paper \cite{gkl}, GKL considered mainly random writing errors: at every time step, with probability $1-\alpha$ the transition follows rules (\ref{gkl-iv-rules})--(\ref{gkl-iv-r}) and with probability $\alpha$ the final state is chosen at random with equal probabilities. In other words, for GKL-IV model under noise level $\alpha$, at every time step the probability of writing the new state to a cell according to the rules is $(1-\alpha)+\frac{1}{4}\alpha$, while the probability of doing it incorrectly is $\frac{3}{4}\alpha$.

A PCA is ergodic if it eventually forgets its initial state, meaning that it has a unique invariant measure---a unique probability distribution of states over the configuration space of the model that does not change under the dynamics \cite{russkiye,mairesse,busic,marcovici,taati17,roberto}. Remarkably, GKL found by means of numerical experiments evidence that GKL-IV may be nonergodic below a certain small level of noise $\alpha^{*} \approx 0.05$ \cite{gkl}. If true, this would provide a counterexample to the positive probabilities conjecture, according to which all one-dimensional PCA with positive rates, short-range interactions and finite local state space are ergodic. This conjecture is deeply rooted in the theory of Markov processes and has a counterpart in the well-known statistical physics lore that one-dimensional systems do not display phase transitions at finite ($T>0$) temperature \cite{holley,bennett85,bennett90,grinstein,cuesta}. It took nearly three decades to disprove this conjecture in general \cite{reliable,gacs,gray}, while counterexamples also appeared in the physics literature
\cite{cuesta,evans,kafri,bjp30,rakos,paessens}. The roles played by the size of the rule spaces, symmetries, number of absorbing states, irreversibility and the thermodynamic limit in the phenomenon are still under debate.

\begin{figure}[t]
\centering
\includegraphics[scale=0.70, clip]{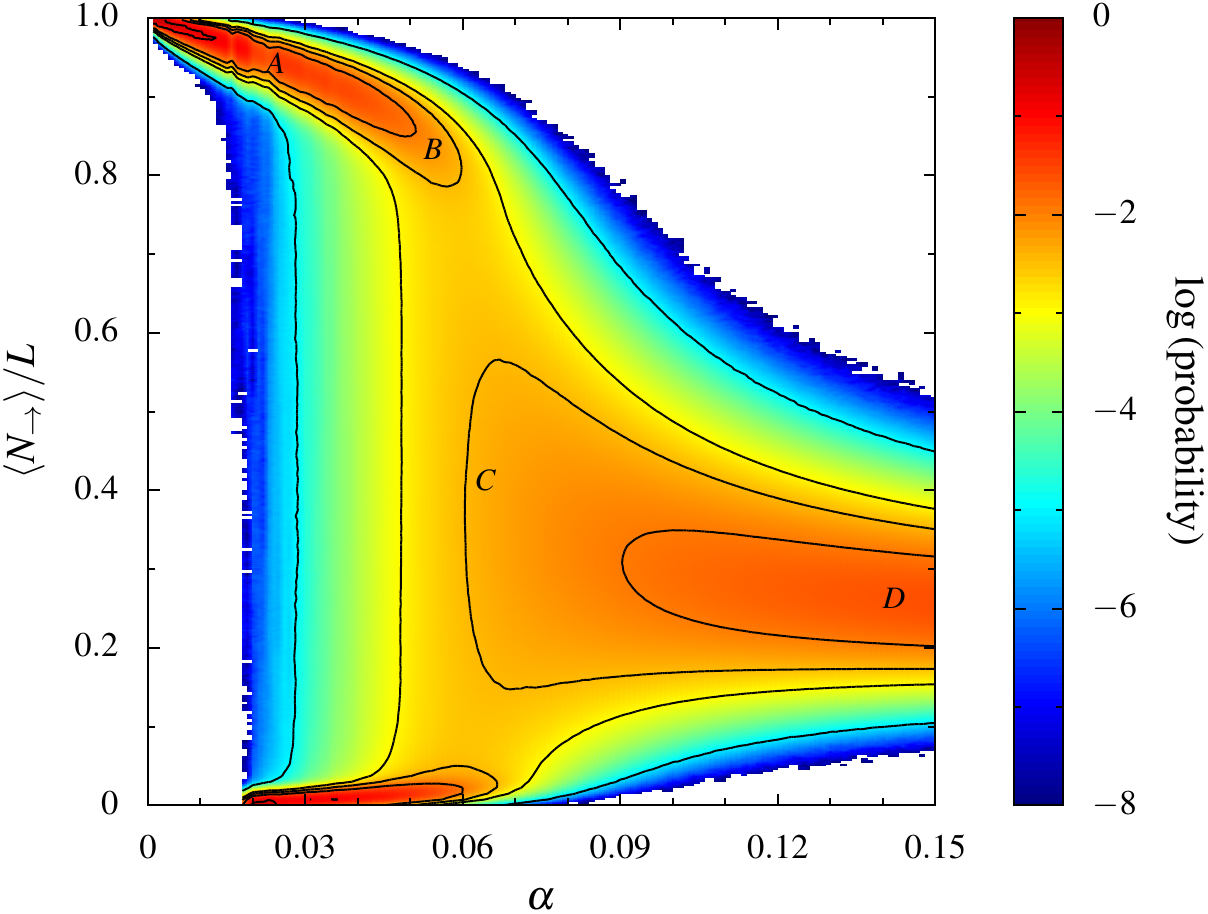}
\caption{\label{fig:heat}Density plot and level curves of the probability density of the majority state in the stationary state of the GKL-IV PCA under noise for an array of $L=400$ cells. For each of the 150 levels of noise $0 < \alpha \leq 0.15$ (in steps of $\Delta\alpha=0.001$), the probability density histogram was obtained from $100$ million samples. The letters mark the points $A$ $(0.012,0.95)$, $B$ $(0.05,0.84)$, $C$ $(0.065,0.40)$, and $D$ $(0.14,0.25)$.}
\end{figure}


\subsection{\label{sec:stationary}Empirical stationary density}

Little is known about the ergodicity of GKL-IV beyond the loose estimate $\alpha^{*} \approx 0.05$ mentioned in \cite{gkl}, in contrast with the same problem for GKL-II \cite{reliable,gacs,gray,maes,assembly}. To improve this situation, we performed relatively large simulations of GKL-IV under noise to verify whether there may be some sort of ergodic-nonergodic transition upon the variation of $\alpha$.

Our simulations ran as follows. For a given level of noise $\alpha$, we initialize a PCA of length $L=400$ with all cells in the state ${\to}$, relax the initial state for $4L$ time steps, and start sampling the number $N_{\to}$ of cells in the state ${\to}$ in the PCA every $5$ time steps.
We choose $4L$ for the initial relaxation time because without noise GKL-IV converges to the absorbing state from an arbitrary initial configuration in $\sim 2L$ time steps maximum, see Figure~\ref{fig:performance}. The choice of $4L$ time steps for relaxation seems reasonable, since we are already starting from an absorbing configuration.
We collected $100$ million samples of the stationary state for each level of noise in the range $0 < \alpha \leq 0.15$ in steps of $\Delta\alpha=0.001$ and the results are displayed as a probability density plot in Figure~\ref{fig:heat}. Each vertical line (fixed $\alpha$) in Figure~\ref{fig:heat} is a histogram of bin size $1/L$ and unit area. Note that the lack of points scattered near $\langle N_{\to} \rangle/L=1$ except for the smallest values of $\alpha$ in Figure~\ref{fig:heat} corroborates \textit{ex post} the choice of $4L$ time steps for the initial relaxation time. In fact, the initial relaxation time is utterly irrelevant---the worst that it can do is to contribute with a couple of hundreds of ``bad samples'' to the set of 100 million samples for each value of $\alpha$.

\begin{figure*}[t]
\centering
\includegraphics[scale=0.85,clip]{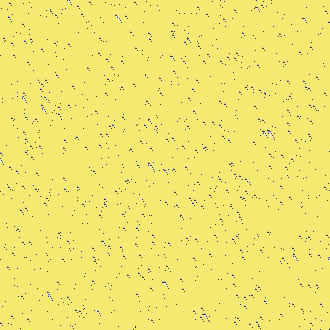} \hspace{1em}
\includegraphics[scale=0.85,clip]{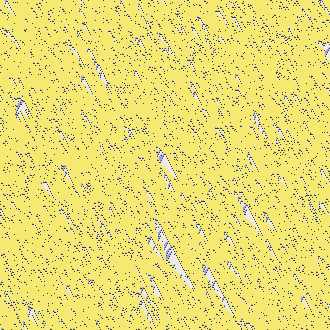} \\[3ex]
\includegraphics[scale=0.85,clip]{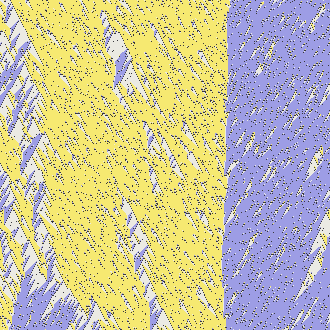} \hspace{1em}
\includegraphics[scale=0.85,clip]{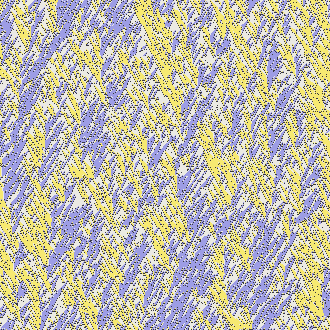}
\caption{\label{fig:runs}Sample runs of GKL-IV ($L=400$ cells) under noise in the stationary state; time flows downward. In left-to-right, top-to-bottom order, the space-time diagrams correspond to typical configurations found around the loci $A$, $B$, $C$ and $D$ indicated in Figure~\ref{fig:heat}. Color coding reads as in Figure~\ref{fig:gkl-iv}, namely, yellow $\equiv {\to}$, purple $\equiv {\leftarrow}$, white $\equiv {\uparrow}$, and black $\equiv {\downarrow}$. Note how white regions of ${\uparrow}$ states tend to cluster, while black regions (mostly just isolated spots) of ${\downarrow}$ states straggle throughout.}
\end{figure*}

Figure~\ref{fig:heat} clearly displays the two extreme behaviors expected of the noisy GKL-IV. When $\alpha=0$, the distribution of $\langle N_{\to} \rangle/L$ is a zero-width distribution concentrated at $1$. At the other extreme, when the level of the noise is high, in our case $\alpha \gtrsim 0.12$, all the states become equiprobable and the density $\langle N_{\to} \rangle/L$ concentrates around $1/4$ with a more or less symmetric distribution that becomes sharper as $\alpha$ increases. The difficult question is whether there is a finite positive $\alpha^{*}>0$ such that the noisy GKL-IV is ergodic above $\alpha^{*}$ and nonergodic below it. Figure~\ref{fig:heat} indicates that the probability distribution of $\langle N_{\to} \rangle/L$ becomes bimodal at $\alpha \approx 0.05$, with one peak concentrated near the majority of states ${\to}$ and the other peak near the majority of states ${\leftarrow}$, becoming narrower as $\alpha \searrow 0$. We also see that flipping between the two majority phases ceases completely at $\alpha \approx 0.016$, at least within the span of $5 \times 10^{8}$ time steps for each given $\alpha$ of our simulations. These seem to indicate that the noisy GKL-IV may be nonergodic for some finite $\alpha \lesssim 0.016$.

Figure~\ref{fig:runs} depicts typical space-time diagrams of the noisy GKL-IV with $L=400$ cells for some selected levels of noise. In diagram $A$ ($\alpha=0.012$, upper left corner), the state of the PCA just fluctuates about the majority state of ${\to}$ to which it would have converged if it were not for the noise. Small islands of contiguous ${\uparrow}$ (white) states that form could in principle foster the spread of the minority state ${\leftarrow}$ (purple), but these islands are too small and short-lived to make any difference. Ergodicity at this level of noise would imply that an island of the minority phase (or of the ${\uparrow}$  (white) state) large enough to thrive in the background of the majority phase and noise can form randomly---an exceedingly unlikely event. The overall result is a spotted spatiotemporal pattern of the majority state that on average occupy $\sim 95\%$ of the cells.

As the noise $\alpha$ increases, larger islands of ${\uparrow}$ (white) states form and the states ${\to}$ (yellow) and ${\leftarrow}$ (purple) tend to coexist for longer periods. In diagram $B$ ($\alpha=0.05$, upper right corner of Figure~\ref{fig:runs}), we see larger islands of ${\uparrow}$ (white) states allowing ${\leftarrow}$ (purple) states to spread to the left until being annihilated. Eventually, however, those sliders meet to form bigger ones, survive for longer periods and become the majority state. Such an event was captured in diagram $C$ ($\alpha=0.065$, lower left corner of Figure~\ref{fig:runs}). Note how the majority of states ${\to}$ (yellow) in the top of the diagram is superseded by the ${\leftarrow}$ (purple) states after some time; at the bottom of diagram $C$ the ${\leftarrow}$ (purple) states occupy $\sim 60\%$ of the cells. The ``border'' at the picture is a remainder of the initial condition; if we follow the evolution of the PCA for a little longer we would see a more homogeneous mixture of states. The PCA has flipped between two majority phases, an indication that at $\alpha=0.065$ it is ergodic.

Finally, under the presence of strong noise, the PCA loses almost all structure except very locally and for short times. This can be seen in diagram $D$ ($\alpha=0.14$, lower right corner of Figure~\ref{fig:runs}). Although islands of the attractive states ${\to}$ (yellow) and ${\leftarrow}$ (purple) endure more than islands of the other two states (and this is particularly true of the ${\downarrow}$ (black) states), on average all four states are present approximately in the same amount. Note, in Figure~\ref{fig:heat}, how the stationary probability density at $C$ still displays a bimodal profile, while at $D$ it is clearly a single-peaked distribution centered at $\sim 1/4$.


\subsection{\label{sec:flipping}Flipping times}

It is possible to qualitatively spot an ergodic phase by the analysis of the flipping times between the different stationary configurations of the model. The idea is that this quantity diverges as ``potential barriers'' grow between the metastable configurations of the system as it gets larger, with the system getting trapped deeper and deeper inside one configuration until ultimately ergodicity is broken in the limit of a system of infinite length. Based on an analogy between the flipping time $\tau(L,\alpha)$ between the majority phases of a PCA of $L$ cells subject to noise level $\alpha$ and the correlation length $\xi_{\|}(L,T)$ of a 2D equilibrium interacting classical spin model of linear size $L$ at temperature $T$ (see \cite{maes,rakos,paessens,assembly,mfrsos} for details), we expect that
\begin{equation}
\label{eq:tau}
\tau(L,\alpha) \sim \exp[u(L,\alpha)].
\end{equation}
A nonergodic dynamics implies that $\tau(L,\alpha)$ diverges as $L \nearrow \infty$, while for an ergodic dynamics $u(L,\alpha)$ remains bounded in $L$, signaling that the PCA forgets about its initial condition in finite time, wandering over the entire configuration space and making the invariant measure unique. Clearly, for GKL-IV $\tau(L,\alpha)$ must diverge at $\alpha=0$. Based on these observations, in a nonergodic phase we must have, to first order, $u(L,\alpha) \sim b(L)/\alpha$ for $\alpha \searrow 0$ and fixed $L$ and $u(L,\alpha) \sim c(\alpha)L$ for fixed $\alpha$ and $L \nearrow \infty$, where $b(L)$ and $c(\alpha)$ are bounded functions of their arguments.

\begin{figure}[t]
\centering
\includegraphics[viewport=10 0 530 430, scale=0.40, clip]{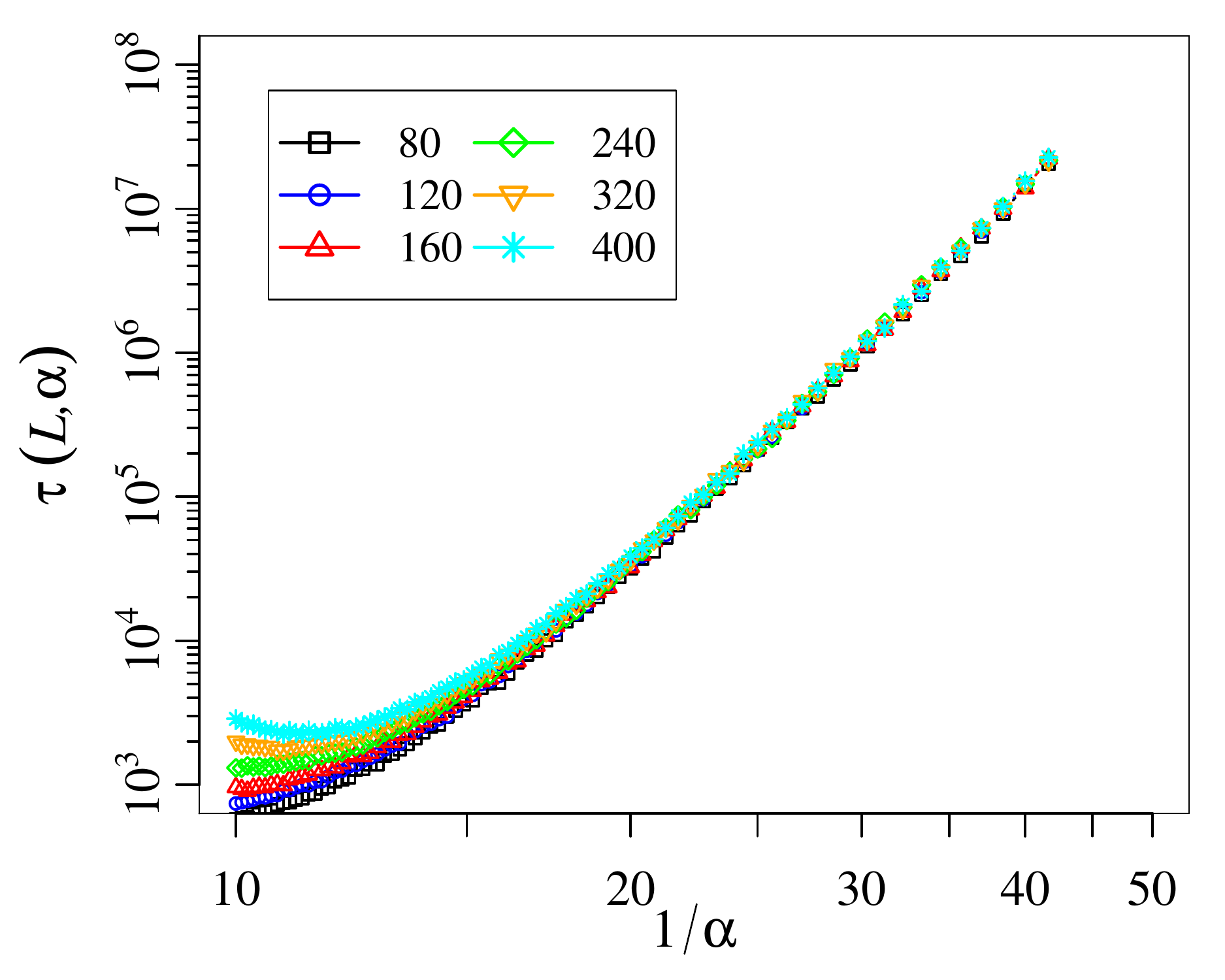}
\caption{\label{fig:flip-a}The flipping times $\tau(L,\alpha)$ ($L$ fixed) grow exponentially as the PCA dynamics becomes less noisy, clearly diverging as $\alpha \searrow 0$.}
\end{figure}

\begin{figure}[t]
\centering
\includegraphics[viewport=0 0 530 430, scale=0.40, clip]{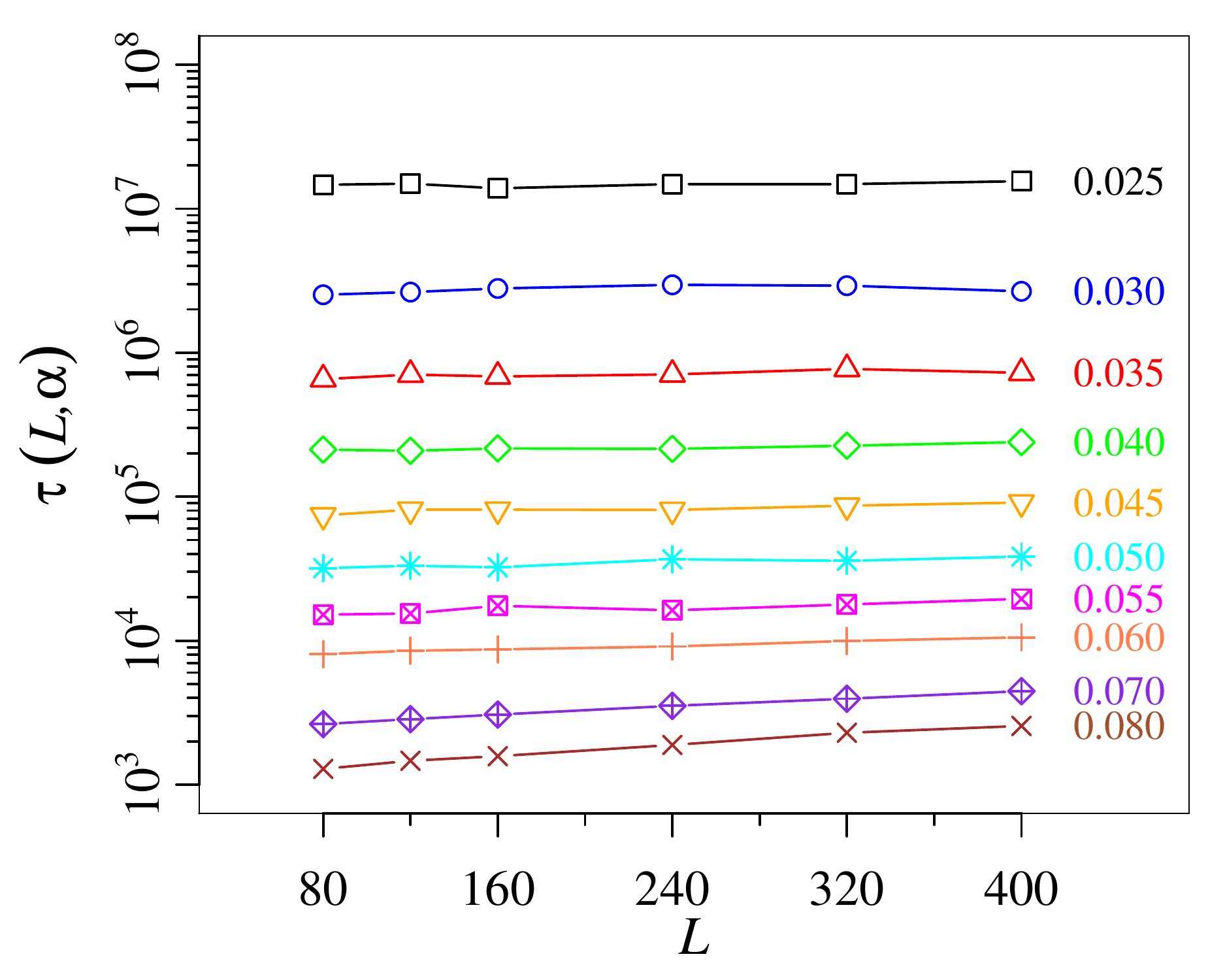}
\caption{\label{fig:flip-l}Behavior of $\tau(L,\alpha)$ with the system size $L$  (fixed $\alpha$). While $\tau(L,\alpha)$ clearly diverges as $\alpha \searrow 0$, it does not so as $L$ grows at least down to $\alpha=0.025$.}
\end{figure}

We measured $\tau(L,\alpha)$ for the noisy GKL-IV as follows. For a given level of noise $\alpha$, we initialize the PCA with all cells in the state ${\to}$ and run the dynamics until a state with majority of cells in the state ${\leftarrow}$ is observed ($N_{\leftarrow}(\tau)>L/2$), signaling that the PCA transposed the ``potential barrier.'' We choose the initial configuration with all cells in the state ${\to}$ because, being an attractive and absorbing state, it provides the ``worst case scenario'' if the PCA has to reach a configuration with a majority of $\leftarrow$. We then obtain the flipping time $\tau(L,\alpha)$ for each given $L$ and $\alpha$ as an average over $1000$ such hitting times. In our simulations $80 \leq L \leq 400$ and $0.024 \leq \alpha \leq 0.100$. We did not take measurements under $\alpha \lesssim 0.02$ because it would take several thousand hours (months, literally) of CPU time on modern workstations to obtain one point. Note that we write low-level C code to run the simulations, and that even the pseudo-random number generator was thought-out to run as fast as possible (we employ Vigna's superb \texttt{xoroshiro128+} generator \cite{xoroshiro}). The relatively small $L$ also allow us to investigate the flipping times without having to wait too much to observe the flips. Our results appear in Figures~\ref{fig:flip-a} and \ref{fig:flip-l}.

The behavior of $\tau(L,\alpha)$ with $\alpha$ seems to indicate that the PCA is nonergodic at least up to $\alpha \lesssim 0.05$. Otherwise, we do not observe any sign of divergence of $\tau(L,\alpha)$ with increasing $L$ up to $L=400$ and down to $\alpha=0.025$, indicating that the PCA is likely to be ergodic in these regions of parameters. The best that we can do with these mixed signals, then, is to combine the bound $\alpha^{*} \leq 0.025$ provided by the behavior of $\tau(L,\alpha)$ with $L$ together with the bound $\alpha^{*} \lesssim 0.016$ provided by Figure~\ref{fig:heat} to set an upper bound on the critical level of noise separating the ergodic from the nonergodic phase of GKL-IV, if any, at $\alpha^{*} \approx 0.016$.


\section{\label{summary}Summary and conclusions}

We found that the GKL-IV model performs well in the density classification problem, with a performance comparable with that of the more well-known model GKL-II. In fact, GKL-IV performs slightly better at the task, even having more states to deal with. The additional states ${\uparrow}$ and ${\downarrow}$ enable GKL-IV to annihilate isolated ${\leftarrow}$ and ${\to}$ states and create local islands of majority states ${\uparrow}$ and ${\downarrow}$ that are then eroded from the boundaries by means of transitions involving the states ${\leftarrow}$ and ${\to}$ that propagate twice as fast as the former processes, thus leading the CA to converge to the majority state among these states. Its somewhat elaborate eroder mechanism turns out to be very effective. Surprisingly, we also found in Section~\ref{upupup} that dilution of the state to be classified by random insertion of ${\uparrow}$ or ${\downarrow}$ states enhances the performance of GKL-IV. This suggests a procedure to boost the performance of the CA in more difficult situations of small imbalance between the number of states ${\leftarrow}$ and ${\to}$: enlarge the CA array (say, by $50\%$) with randomly inserted ${\uparrow}$ arrows, at the cost of increased time to complete the classification. On the negative side, GKL-IV takes longer (almost $3$ times more, see Figure~\ref{fig:performance}) to reach consensus. If performance is to be preferred over speed, however, than GKL-IV is a better classifier that GKL-II at only a moderate increase in runtime.

We also investigated the performance of GKL-IV under the influence of noise and found signs of an ergodic-nonergodic phase transition at some small finite positive level of noise. Indeed, from Figure~\ref{fig:heat} we see that the stationary density of ${\to}$ states clearly becomes bimodal below $\alpha \approx 0.05$, indicating that GKL-IV apparently becomes nonergodic for levels of noise below this value, but the exact location of $\alpha^{*}$ is not very clear from that figure. The behavior of the flipping times $\tau(L,\alpha)$ displayed in Figures~\ref{fig:flip-a} and \ref{fig:flip-l} also indicate that GKL-IV is probably nonergodic for $\alpha < 0.05$, although the behavior of this quantity with the system size $L$ indicate that the system is ergodic at least down to $\alpha \approx 0.025$. Combining these somewhat conflicting informations together with the fact that flipping between the two majority phases ceases completely about $\alpha \approx 0.016$, the best that we can do is to set $\alpha^{*} \lesssim 0.016$ as an upper bound on the critical point for a putative ergodic-nonergodic phase transition of GKL-IV. Note that estimates of $\alpha^{*}$ from Figures~\ref{fig:heat}, \ref{fig:flip-a} and \ref{fig:flip-l} are affected by the finite size of the system and the finite time of the simulations. Our data indicate that the noisy GKL-IV may be nonergodic but are not conclusive; larger systems simulated for longer periods could tell better.

In \cite{gkl} the authors advanced the idea that the noisy GKL-IV (as well as its siblings GKL-II and GKL-VI) is ``quasinonergodic,'' in the sense that while the models are ergodic for any $\alpha>0$, convergence to the unique invariant measure is extremely slow. The behavior displayed by $\langle N_{\to} \rangle/L$ and $\tau(L,\alpha)$ in Figures~\ref{fig:heat}, \ref{fig:flip-a} and \ref{fig:flip-l} supports this idea. It should be remarked that while the ergodicity of one-dimensional deterministic CA is in general undecidable, most PCA are believed to be ergodic, with the notable exception of Gacs' very complicated (and still controversial) counterexample \cite{reliable,gacs,gray,mairesse,busic,marcovici,taati17,roberto}. We established an upper bound on the critical level of noise of GKL-IV above which it becomes ergodic. Whether this critical level is smaller or zero remains an open question.

The GKL-II and, principally, GKL-IV CA and PCA deserve more analytical studies. We believe that already at the level of single-cell mean field approximation \cite{dickman,bollobas} the equations may reveal an interesting structure. In the same vein, a study of the spreading of damage \cite{kohring,tome} in the GKL-II and IV PCA may help to clarify the rate of convergence of the dynamics to the stationary states and help to understand CA and PCA that are able to classify density.


\begin{acknowledgments}

We thank Peter Gacs (BU) for useful correspondence, library specialist Angela K. Gruendl (UIUC) for excellent service in providing a copy of reference~\cite{dobrushin}, and Yeva Gevorgyan (USP) for help with the literature in Russian. This work was partially supported by FAPESP (Brazil) under grant nr.~2015/21580-0 (JRGM) and CNPq (Brazil) through Ph.\,D. grant nr.~140684/2016-6 (REOS).


\end{acknowledgments}


\appendix*

\section{\label{ruletable}Complete GKL-IV rule table}

We had to tinker a bit with GKL-IV before getting its rule table right, so we share the result of our labor here. Table \ref{tab:gkl_iv_rules} displays all the elementary transitions of GKL-IV according to rules (\ref{gkl-iv-1})--(\ref{gkl-iv-3}) supplemented by their reflections (\ref{gkl-iv-r}) as described in Section~\ref{model}.

\begin{table*}[t]
\caption{\label{tab:gkl_iv_rules}GKL-IV rule table. When a transition is defined by more than one rule, only the first rule that defines it is listed. Transitions defined by a reflected rule have the respective rule marked by an asterisk.}
\centering
\begin{tabular}{|ccl|ccl|ccl|ccl|ccl|ccl|ccl|ccl|}
\hline\hline
$(x_{i-1},\; x_{i},\; x_{i+1})$ & $x_{i}'$ & Rule &
$(x_{i-1},\; x_{i},\; x_{i+1})$ & $x_{i}'$ & Rule &
$(x_{i-1},\; x_{i},\; x_{i+1})$ & $x_{i}'$ & Rule &
$(x_{i-1},\; x_{i},\; x_{i+1})$ & $x_{i}'$ & Rule \\
\hline
$({\to},{\to},{\to})$ & ${\to}$ & (\ref{gkl-iv-1}) &
$({\to},{\uparrow},{\to})$ & ${\to}$ & (\ref{gkl-iv-1}) &
$({\to},{\leftarrow},{\to})$ & ${\downarrow}$ & (\ref{gkl-iv-2})$^{*}$ &
$({\to},{\downarrow},{\to})$ & ${\to}$ & (\ref{gkl-iv-1}) \\
$({\to},{\to},{\uparrow})$ & ${\to}$ & (\ref{gkl-iv-1}) &
$({\to},{\uparrow},{\uparrow})$ & ${\to}$ & (\ref{gkl-iv-1}) &
$({\to},{\leftarrow},{\uparrow})$ & ${\downarrow}$ & (\ref{gkl-iv-2})$^{*}$ &
$({\to},{\downarrow},{\uparrow})$ & ${\to}$ & (\ref{gkl-iv-1}) \\
$({\to},{\to},{\leftarrow})$ & ${\to}$ & (\ref{gkl-iv-2}) &
$({\to},{\uparrow},{\leftarrow})$ & ${\uparrow}$ & (\ref{gkl-iv-3}) &
$({\to},{\leftarrow},{\leftarrow})$ & ${\leftarrow}$ & (\ref{gkl-iv-2})$^{*}$ &
$({\to},{\downarrow},{\leftarrow})$ & ${\uparrow}$ & (\ref{gkl-iv-3}) \\
$({\to},{\to},{\downarrow})$ & ${\to}$ & (\ref{gkl-iv-1}) &
$({\to},{\uparrow},{\downarrow})$ & ${\to}$ & (\ref{gkl-iv-1}) &
$({\to},{\leftarrow},{\downarrow})$ & ${\leftarrow}$ & (\ref{gkl-iv-2})$^{*}$ &
$({\to},{\downarrow},{\downarrow})$ & ${\to}$ & (\ref{gkl-iv-1}) \\
\hline
$({\uparrow},{\to},{\to})$ & ${\downarrow}$ & (\ref{gkl-iv-2}) &
$({\uparrow},{\uparrow},{\to})$ & ${\uparrow}$ & (\ref{gkl-iv-3}) &
$({\uparrow},{\leftarrow},{\to})$ & ${\downarrow}$ & (\ref{gkl-iv-2})$^{*}$ &
$({\uparrow},{\downarrow},{\to})$ & ${\uparrow}$ & (\ref{gkl-iv-3}) \\
$({\uparrow},{\to},{\uparrow})$ & ${\downarrow}$ & (\ref{gkl-iv-2}) &
$({\uparrow},{\uparrow},{\uparrow})$ & ${\uparrow}$ & (\ref{gkl-iv-3}) &
$({\uparrow},{\leftarrow},{\uparrow})$ & ${\downarrow}$ & (\ref{gkl-iv-2})$^{*}$ &
$({\uparrow},{\downarrow},{\uparrow})$ & ${\uparrow}$ & (\ref{gkl-iv-3}) \\
$({\uparrow},{\to},{\leftarrow})$ & ${\downarrow}$ & (\ref{gkl-iv-2}) &
$({\uparrow},{\uparrow},{\leftarrow})$ & ${\leftarrow}$ & (\ref{gkl-iv-1})$^{*}$ &
$({\uparrow},{\leftarrow},{\leftarrow})$ & ${\leftarrow}$ & (\ref{gkl-iv-1})$^{*}$ &
$({\uparrow},{\downarrow},{\leftarrow})$ & ${\leftarrow}$ & (\ref{gkl-iv-1})$^{*}$ \\
$({\uparrow},{\to},{\downarrow})$ & ${\downarrow}$ & (\ref{gkl-iv-2}) &
$({\uparrow},{\uparrow},{\downarrow})$ & ${\uparrow}$ & (\ref{gkl-iv-3}) &
$({\uparrow},{\leftarrow},{\downarrow})$ & ${\leftarrow}$ & (\ref{gkl-iv-2})$^{*}$ &
$({\uparrow},{\downarrow},{\downarrow})$ & ${\uparrow}$ & (\ref{gkl-iv-3}) \\
\hline
$({\leftarrow},{\to},{\to})$ & ${\downarrow}$ & (\ref{gkl-iv-2}) &
$({\leftarrow},{\uparrow},{\to})$ & ${\uparrow}$ & (\ref{gkl-iv-3}) &
$({\leftarrow},{\leftarrow},{\to})$ & ${\downarrow}$ & (\ref{gkl-iv-2})$^{*}$ &
$({\leftarrow},{\downarrow},{\to})$ & ${\uparrow}$ & (\ref{gkl-iv-3}) \\
$({\leftarrow},{\to},{\uparrow})$ & ${\downarrow}$ & (\ref{gkl-iv-2}) &
$({\leftarrow},{\uparrow},{\uparrow})$ & ${\uparrow}$ & (\ref{gkl-iv-3}) &
$({\leftarrow},{\leftarrow},{\uparrow})$ & ${\downarrow}$ & (\ref{gkl-iv-2})$^{*}$ &
$({\leftarrow},{\downarrow},{\uparrow})$ & ${\uparrow}$ & (\ref{gkl-iv-3}) \\
$({\leftarrow},{\to},{\leftarrow})$ & ${\downarrow}$ & (\ref{gkl-iv-2}) &
$({\leftarrow},{\uparrow},{\leftarrow})$ & ${\leftarrow}$ & (\ref{gkl-iv-1})$^{*}$ &
$({\leftarrow},{\leftarrow},{\leftarrow})$ & ${\leftarrow}$ & (\ref{gkl-iv-1})$^{*}$ &
$({\leftarrow},{\downarrow},{\leftarrow})$ & ${\leftarrow}$ & (\ref{gkl-iv-1})$^{*}$ \\
$({\leftarrow},{\to},{\downarrow})$ &  ${\downarrow}$ & (\ref{gkl-iv-2}) &
$({\leftarrow},{\uparrow},{\downarrow})$ &  ${\uparrow}$ & (\ref{gkl-iv-3}) &
$({\leftarrow},{\leftarrow},{\downarrow})$ &  ${\leftarrow}$ & (\ref{gkl-iv-2})$^{*}$ &
$({\leftarrow},{\downarrow},{\downarrow})$ &  ${\uparrow}$ & (\ref{gkl-iv-3}) \\
\hline
$({\downarrow},{\to},{\to})$ & ${\to}$ & (\ref{gkl-iv-2}) &
$({\downarrow},{\uparrow},{\to})$ & ${\uparrow}$ & (\ref{gkl-iv-3}) &
$({\downarrow},{\leftarrow},{\to})$ & ${\downarrow}$ & (\ref{gkl-iv-2})$^{*}$ &
$({\downarrow},{\downarrow},{\to})$ & ${\uparrow}$ & (\ref{gkl-iv-3}) \\
$({\downarrow},{\to},{\uparrow})$ & ${\to}$ & (\ref{gkl-iv-2}) &
$({\downarrow},{\uparrow},{\uparrow})$ & ${\uparrow}$ & (\ref{gkl-iv-3}) &
$({\downarrow},{\leftarrow},{\uparrow})$ & ${\downarrow}$ & (\ref{gkl-iv-2})$^{*}$ &
$({\downarrow},{\downarrow},{\uparrow})$ & ${\uparrow}$ & (\ref{gkl-iv-3}) \\
$({\downarrow},{\to},{\leftarrow})$ & ${\to}$ & (\ref{gkl-iv-2}) &
$({\downarrow},{\uparrow},{\leftarrow})$ & ${\leftarrow}$ & (\ref{gkl-iv-1})$^{*}$ &
$({\downarrow},{\leftarrow},{\leftarrow})$ & ${\leftarrow}$ & (\ref{gkl-iv-1})$^{*}$ &
$({\downarrow},{\downarrow},{\leftarrow})$ & ${\leftarrow}$ & (\ref{gkl-iv-1})$^{*}$ \\
$({\downarrow},{\to},{\downarrow})$ & ${\to}$ & (\ref{gkl-iv-2}) &
$({\downarrow},{\uparrow},{\downarrow})$ & ${\uparrow}$ & (\ref{gkl-iv-3}) &
$({\downarrow},{\leftarrow},{\downarrow})$ & ${\leftarrow}$ & (\ref{gkl-iv-2})$^{*}$ &
$({\downarrow},{\downarrow},{\downarrow})$ & ${\uparrow}$ & (\ref{gkl-iv-3}) \\
\hline\hline
\end{tabular}
\end{table*}


\end{document}